\documentclass[conference]{IEEEtran}
\usepackage[T1]{fontenc}
\usepackage{url}
\usepackage{cite}
\usepackage{amsmath,amssymb,amsfonts}
\usepackage{algorithm}
\usepackage{algorithmic}
\usepackage{graphicx}
\usepackage{textcomp}
\usepackage{xcolor}

\usepackage{booktabs}
\usepackage{multirow}
\usepackage{threeparttable}
\usepackage[caption=false,font=footnotesize]{subfig}
\def\BibTeX{{\rm B\kern-.05em{\sc i\kern-.025em b}\kern-.08em
    T\kern-.1667em\lower.7ex\hbox{E}\kern-.125emX}}
    
\begin{document}

\title{QCOEM: Quantum Cloud Orchestration with Evolutionary Multi-Objective Optimization}


\author{
    \IEEEauthorblockN{Tam N. Pham\textsuperscript{1,2}, Hoa T. Nguyen\textsuperscript{3}, and Quan Le-Trung\textsuperscript{1,2}}
    \IEEEauthorblockA{\textsuperscript{1}\textit{University of Information Technology, Ho Chi Minh City 700000, Vietnam}}
    \IEEEauthorblockA{\textsuperscript{2}\textit{Vietnam National University, Ho Chi Minh City 700000, Vietnam}}
    \IEEEauthorblockA{\textsuperscript{3}\textit{Data61, CSIRO, Clayton, 3168, Victoria, Australia}} 
    tampn.18@grad.uit.edu.vn, hoa.nguyen@csiro.au, quanlt@uit.edu.vn
}


\maketitle

\begin{abstract}
Quantum cloud platforms need to dynamically orchestrate workloads across heterogeneous quantum computation backends whose noise profiles, qubit topologies, and queues vary over time. Existing orchestrators use noise-agnostic heuristics that ignore backend-specific errors, causing reduced execution fidelity, load imbalance, and frequent rescheduling. To address these challenges, we propose QCOEM - a Quantum Cloud Orchestration framework that leverages Evolutionary algorithms for Multi-objective optimization of quantum task scheduling. We compare NSGA-II and NSGA-III for jointly minimizing mean completion time, execution error rate, and load imbalance. To select schedules from a non-convex Pareto front, we apply an Augmented Achievement Scalarization Function (AASF) as a preference-based decision rule that maps the Pareto set to a single dispatchable schedule aligned with user priorities. Our extensive performance evaluation in a heterogeneous quantum cloud environment shows zero task rescheduling and about 30\% higher mean fidelity than noise-agnostic heuristics, while maintaining bounded scheduling overhead. The experiment results indicate that our QCOEM framework can deliver stable, high-fidelity execution and lightweight resource management for quantum cloud computing.
\end{abstract}


\section{Introduction}
The practical deployment of quantum computing has progressed from primarily theoretical investigations to large-scale cloud-based services, leading to the emergence of Quantum-as-a-Service (QaaS) platforms \cite{ravi2021adaptive, nguyen_quantumcloudcomputing_2024}. However, existing quantum hardware still remains in the Noisy Intermediate-Scale Quantum (NISQ) era \cite{preskill2018quantum}. At this stage, physical Quantum Processing Units (QPUs) expose only a limited number of qubits and are subject to gate errors, readout errors, short coherence times, and restricted qubit connectivity. These characteristics also vary across devices and drift over time, so identical circuits can achieve markedly different fidelities depending on the backend that executes them \cite{10.1145/3712285.3759785}.

To exploit NISQ hardware, the hybrid quantum-classical execution model, in which quantum circuits run alongside classical processing, has become the dominant approach. Serving such workloads at scale motivates cloud-native platforms that integrate heterogeneous QPUs into orchestration systems such as Kubernetes \cite{stirbu2024qubernetes, faro2023middleware}. Existing platforms, however, primarily focus on executing hybrid workloads and offer limited support for noise-aware orchestration. In particular, efficiently routing quantum circuits to suitable backends based on their device-level error characteristics remains largely an open problem. 

Conventional cloud orchestrators rely on classical heuristics such as Greedy and Round Robin that are tailored to maximize throughput and reduce latency \cite{li2025qusplit, 10.1145/3712285.3759785, madni2017performance, soltani2017heuristic}. However, applying these noise-agnostic strategies to quantum workloads can be systematically suboptimal. For example, dispatching a deep and complex circuit to an idle but error-prone QPU lowers fidelity and may force costly re-execution, whereas repeatedly favouring a few high-quality QPUs creates queueing hotspots that can degrade fairness. Thus, quantum resource orchestration is inherently a Multi-Objective Optimization Problem (MOOP). An effective scheduler must carefully navigate three competing goals: system efficiency, computational accuracy, and operational stability \cite{10.5555/3767901.3767925}. The difficulty is compounded by per-operation errors that accumulate multiplicatively with circuit depth \cite{10.1287/ijoc.2024.0551, 10.1145/3297858.3304075}, resulting in a non-convex Pareto front. Consequently, techniques such as weighted-sum scalarization can recover only solutions on the convex hull of the front and cannot reach optimal policies in its concave regions \cite{deb2013evolutionary}.

To address these challenges, we propose QCOEM, a \textbf{Q}uantum \textbf{C}loud \textbf{O}rchestration framework that leverages different \textbf{E}volutionary optimization algorithms for \textbf{M}ulti-objective quantum task scheduling. Rather than relying on idealized analytical models, we utilize quantum backend (QNode) architecture that is calibrated using hardware snapshot data (such as coupling maps, gate durations, and error rates) obtained from real devices, thereby constructing a physics-informed evaluation setting. In our experiments, these snapshots are assumed to remain stable across runs, while the orchestration loop maintains awareness of queue conditions via heartbeat telemetry. By integrating this environment with Multi-Objective Evolutionary Algorithms (MOEAs), the framework can effectively navigate and operate within the current limitations of heterogeneous QPUs.

The primary contributions of this paper are as follows:

\begin{itemize}
    \item We propose QCOEM as a noise- and delay-aware quantum orchestration framework that leverages calibrated hardware snapshots and heartbeat-based queueing-delay estimates to jointly capture mean completion time, mean error rate, and load imbalance.

    \item We leverage the NSGA-II and NSGA-III approaches to approximate a non-dominated set of task-to-QNode assignments and use an Augmented Achievement Scalarization Function (A    ASF) to perform preference-based selection from this candidate set. To our knowledge, this is one of the first quantum cloud scheduling frameworks to combine AASF-based decision making with an explicit load-imbalance objective.

    \item We evaluate QCOEM on a heterogeneous quantum cloud environment using MQTBench workloads, demonstrating trade-offs among latency, fidelity, and stability, achieving zero rescheduling and more than $30\%$ higher mean fidelity than noise-agnostic baselines, while maintaining bounded scheduling overhead.
\end{itemize}

The remainder of this paper is organized as follows: Section~II details the system model and mathematical formulation. Section~III presents the proposed evolutionary orchestration framework. Section~IV discusses the empirical performance evaluation. Section~V reviews related work, and Section~VI concludes the paper.

\section{System Model and Methodology}

\subsection{Problem Formulation}
Let $\mathcal{T} = \{ \tau_1, \tau_2, \dots, \tau_M \}$ be the set of $M$ incoming quantum tasks, and $\mathcal{Q} = \{ q_1, q_2, \dots, q_N \}$ be the set of $N$ available quantum nodes (QNodes). We define a binary decision variable $x_{ij} \in \{0, 1\}$, where $x_{ij}=1$ if task $\tau_i$ is assigned to QNode $q_j$. The quantum task orchestration is formulated as a Multi-Objective Optimization Problem (MOOP) of minimising the objective vector $\mathbf{F}(\mathbf{X}) =[f_1, f_2, f_3]$ under hardware constraints.

\subsubsection{Mean Completion Time ($f_1$)}
The execution duration of a quantum circuit depends on its complexity and physical gate latency. Following the timing model in the existing studies \cite{nguyen2025qfor, 10.1145/3712285.3759785}, we  estimate the circuit execution time $t_{exec}$ based on the Critical Path of the circuit's Directed Acyclic Graph (DAG):
\begin{equation}
    t_{exec}(\tau_i, q_j) = \left( \sum_{g \in \mathcal{P}_i} t_{g,j} \right) \times S_i
\end{equation}
where $\mathcal{P}_i$ is the set of gates on the critical path of task $\tau_i$, $t_{g,j}$ is the physical gate length on QNode $q_j$, and $S_i$ is the number of execution shots. We assume that all gates on the critical path execute sequentially and that shots are executed sequentially rather than in batches.

The first objective captures delay-aware system efficiency by minimizing the \emph{mean completion time} of each QNode, computed from its current queueing delay (waiting-time estimate) and the estimated execution time of the tasks assigned to it. Let $H_j$ denote the current waiting-time estimate for QNode $q_j$ at the start of the scheduling round, computed from the QNode's next available time $t^{(j)}_{\text{next}}$ and the scheduler's current time $t_{\text{now}}$ as:
\begin{equation}
\label{eq:hj}
    H_j = \max\{0,\ t^{(j)}_{\text{next}} - t_{\text{now}}\}
\end{equation}
\noindent The total completion time at QNode $q_j$ is denoted by $L_j(\mathbf{X})$ and is given by:
\begin{equation}
\label{eq:Lj}
    L_j(\mathbf{X}) = H_j + \sum_{i=1}^{M} x_{ij} \cdot t_{exec}(\tau_i, q_j)
\end{equation}
The objective minimizes the mean completion time across all QNodes:
\begin{equation}
    f_1(\mathbf{X}) = \frac{1}{N} \sum_{j=1}^{N} L_j(\mathbf{X})
\end{equation}

\subsubsection{Mean Error Rate ($f_2$)}
The reliability of NISQ devices decreases significantly with circuit depth and complexity. We estimate the fidelity of task $\tau_i$ on QNode $q_j$ as the product of per-operation success probabilities:
\begin{equation}
    \mathcal{F}(\tau_i, q_j) = \prod_{g \in \mathcal{G}_i} \left(1-\epsilon_{g,j}\right)
\end{equation}
where $\mathcal{G}_i$ is the set of all operations in $\tau_i$ (including readout), and $\epsilon_{g,j}$ is the calibrated gate and readout error rate on QNode $q_j$. The corresponding error rate per-task is $1-\mathcal{F}(\tau_i,q_j)$, and the objective minimizes the mean error rate over all tasks in the batch:
\begin{equation}
    f_2(\mathbf{X}) = \frac{1}{M} \sum_{i=1}^{M} \sum_{j=1}^{N} x_{ij} \cdot \left(1-\mathcal{F}(\tau_i, q_j)\right)
\end{equation}

\subsubsection{Load Imbalance ($f_3$)}
To discourage queueing hotspots and promote fair utilization across the available QNodes, we apply spatial load balancing by minimizing the spread of QNode completion times. The third objective is therefore given by the standard deviation of $L_j(\mathbf{X})$, where $f_1(\mathbf{X})$ denotes the mean QNode completion time across all QNodes at evaluation time.
\begin{equation}
\begin{aligned}
    f_3(\mathbf{X}) &= \sqrt{\frac{1}{N} \sum_{j=1}^{N} \left( L_j(\mathbf{X}) - f_1(\mathbf{X}) \right)^2}
\end{aligned}
\end{equation}

\subsubsection{Constraints}
The orchestration is subject to the following constraints. 

\noindent 1) Indivisible task ($C_1$): Each task is executed as a single quantum circuit without splitting or partitioning.

\begin{equation}\label{eq_constr:1}
    C_1: \; \mathrm{Size}(\tau_i) = 1, \quad \forall \tau_i \in \mathcal{T}
\end{equation}

\noindent 2) Unique assignment ($C_2$): Each task is assigned to exactly one QNode.

\begin{equation}\label{eq_constr:2}
    C_2: \; \sum_{j=1}^{N} x_{ij} = 1, \quad \forall i \in \{1, \dots, M\}
\end{equation}

\noindent 3) Qubit capacity ($C_3$): A task can be mapped to a QNode only if sufficient physical qubits are available.

\begin{equation}\label{eq_constr:3}
    C_3: \;\tau_i^{r} \le q_j^{c}, \quad \forall i \in \{1, \dots, M\},\ \forall j \in \{1, \dots, N\}
\end{equation}
where $\tau_i^{r}$ represents the qubit requirement of task $\tau_i$, and $q_j^{c}$ denotes the physical qubit capacity of QNode $q_j$.

\subsubsection{Problem Formulation}
In each orchestration round, a batch of $M$ ready tasks $\mathcal{T}$ becomes available, together with a pool of $N$ QNodes $\mathcal{Q}$. Every QNode $q_j$ reports its current waiting-time estimate $H_j$ (derived from its queue state) and its calibrated timing and noise parameters, which are used to compute $f_1$, $f_2$, and $f_3$. Each task must be scheduled as an indivisible circuit and mapped to exactly one QNode that can feasibly execute it ($C_1$--$C_3$). User priorities are encoded in a weight vector $\mathbf{w} = [w_1, w_2, w_3]$.

The orchestrator need to determine a feasible task-to-QNode assignment $\mathbf{X}$ that minimizes:
\begin{equation}
\label{eq:moop}
\min_{\mathbf{X}}\ \mathbf{F}(\mathbf{X})=[f_1(\mathbf{X}),\ f_2(\mathbf{X}),\ f_3(\mathbf{X})] \quad \text{s.t. } C_1\text{-}C_3
\end{equation}

\noindent where $\mathbf{X}=[x_{ij}]$ with $x_{ij}=1$ if task $\tau_i$ is assigned to QNode $q_j$, and $f_1$, $f_2$, and $f_3$ are the mean completion time, mean error rate and load-balance objectives defined above.

\subsection{Evolutionary Optimization and Scalarization Strategy}
With a search space of $N^M$, evaluating all task-to-QNode mappings is an NP-hard combinatorial problem. We employ the Non-dominated Sorting Genetic Algorithm II/III (NSGA-II, NSGA-III) \cite{deb2002fast, deb2013evolutionary, jain2013evolutionary, deb2011multi} to heuristically approximate the Pareto-optimal front. Because quantum noise is heterogeneous and tasks are discrete, the objective space is inherently non-convex. As a result, classical weighted-sum decision-making has a duality gap and cannot find optimal schedules in concave regions of the Pareto front.

Before scalarization, we normalize each objective to a comparable scale using min-max normalization:
\begin{equation}
\label{eq:norm}
    f'_k(\mathbf{X}) = \frac{f_k(\mathbf{X}) - f_k^{\min}}{f_k^{\max} - f_k^{\min}}
\end{equation}
\noindent where $f_k^{\min}$ and $f_k^{\max}$ are the minimum and maximum values of the $k$-th objective observed among the candidate solutions. This step prevents any objective with a larger numeric range from dominating the scalarization and improves the interpretability of user preferences $\mathbf{w}$.

To resolve the non-convexity issue, the proposed framework utilizes the Augmented Achievement Scalarization Function (AASF) for post-optimization selection \cite{wierzbicki1982mathematical, singh2020investigating}.
\begin{equation}
    \min_{\mathbf{X}} S(\mathbf{X}) = \max_{k \in \{1,2,3\}} \left[ \frac{f'_k(\mathbf{X}) - z^*_k}{w_k} \right] + \rho \sum_{k=1}^{3} \frac{f'_k(\mathbf{X}) - z^*_k}{w_k}
    \label{eq:aasf}
\end{equation}
where $f'_k$ is the normalized objective value, $z_k^*$ is the ideal reference point, $w_k$ denotes the user-defined weight preference, and $\rho$ is a small enhancement parameter (e.g., $10^{-4}$) used to augment the achievement function. We use AASF as a scalarizing decision function by computing $S(\mathbf{X})$ for each MOEA-generated candidate (typically non-dominated) schedule and selecting $\arg\min S(\mathbf{X})$ under user preference weights, under scalarization assumptions.

\section{QCOEM Framework}
\label{sec:framework}


\begin{figure}[t]
    \centering
    \includegraphics[width=0.8\linewidth]{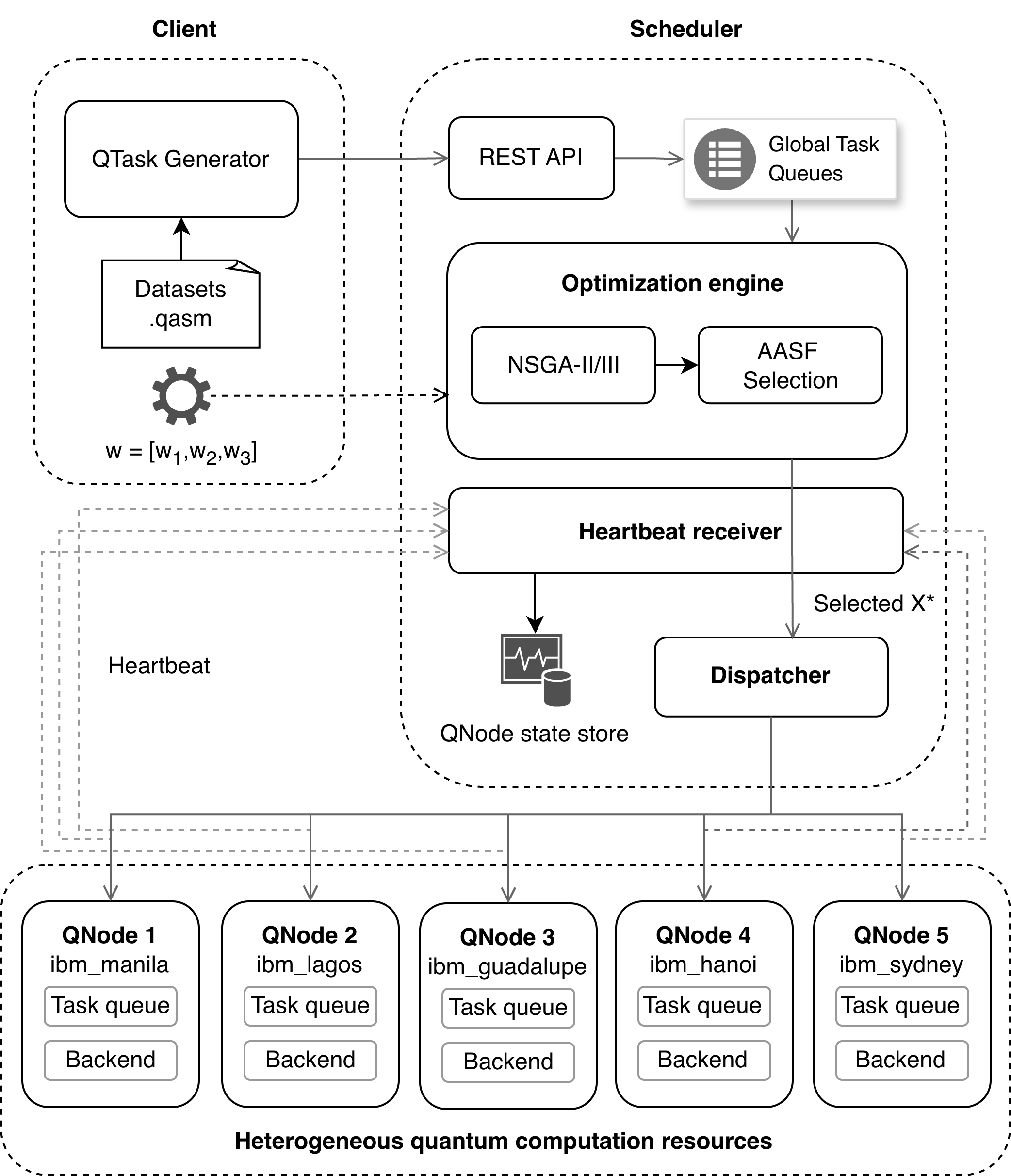}
    \caption{System overview of the proposed QCOEM framework. Periodic QNode heartbeats asynchronously update the scheduler's cached state; each orchestration round uses this state to compute a batch assignment, which is then dispatched to QNodes for FIFO execution.}
    \label{fig:system_overview}
\end{figure}

\subsection{System Architecture}
As illustrated in Fig.~\ref{fig:system_overview}, QCOEM is deployed on Kubernetes with a centralized scheduler and a pool of QNodes; heartbeats provide queue and calibration state, and the scheduler computes and dispatches batch assignments.

\subsubsection{Scheduler} is the control plane that batches tasks, optimizes assignments, and dispatches them. It includes:
\begin{itemize}
    \item Global queue: Receives tasks and forms batches of size $M$ for NSGA-II/NSGA-III.
    \item Optimizer: Estimates execution time/fidelity per task-to-QNode pair from transpilation and calibration data, solves the MOOP, and selects one schedule via AASF under weights $\mathbf{w}$.
    \item Heartbeat receiver: Maintains per-QNode availability, calibration snapshot, and queue estimates (e.g., next available time).
    \item Dispatcher: Sends $\mathbf{X}^*$ to QNodes and updates queue state.
\end{itemize}

\subsubsection{QNodes}
Each QNode is a data-plane worker that couples a calibrated backend profile with lightweight telemetry and sequential circuit execution. The system exposes $N$ QNodes as network-accessible services; each QNode emulates a specific quantum backend using a calibration snapshot (e.g., $T_1$, $T_2$, gate/readout error rates, and coupling map) and executes requests via a local, non-preemptive first-in, first-out (FIFO) queue. Each QNode comprises:

\begin{itemize}
    \item Quantum backend emulator: a calibrated Qiskit \texttt{FakeBackendV2}\footnote{\url{https://quantum.cloud.ibm.com/docs/en/api/qiskit-ibm-runtime/fake-provider}} snapshot that captures noise, gate durations, and connectivity constraints.

    \item Task executor: parses incoming QASM circuits, transpiles them to the QNode's native gate set, and executes circuits sequentially under the snapshot-derived noise model.

    \item Heartbeat: periodically reports availability and queue state (such as next available time and number of pending tasks), enabling the Scheduler to maintain an up-to-date cluster view.
\end{itemize}

\subsection{Scheduling Workflow}
The scheduler operates in discrete orchestration rounds as illustrated in Algorithm~\ref{alg:framework}. In each round, it ingests QNode heartbeats to estimate availability and queueing delay, applies NSGA-II/NSGA-III to approximate the Pareto set for the batch assignment, then selects a single schedule via AASF under preference weights $\mathbf{w}$ and dispatches tasks to QNodes.
    
\begin{algorithm}[htbp]
\caption{QCOEM Scheduling Workflow}
\label{alg:framework}
\begin{algorithmic}[1]
\REQUIRE Batch of tasks $\mathcal{T}$; QNodes $\mathcal{Q}$; preference weights $\mathbf{w}$; MOEA configs ($P$, $G_{\max}$); termination criteria (ftol, xtol, cvtol, window)
\ENSURE Selected assignment matrix $\mathbf{X}^*$
\STATE Ingest heartbeat and compute waiting-time estimate $H_j$ for all $q_j \in \mathcal{Q}$
\STATE Initialize population $\mathcal{A}$ of $P$ valid assignment vectors
\STATE $g \leftarrow 1$
\WHILE{$g \le G_{\max}$ \AND termination criteria not met}
    \STATE Generate offspring $\mathcal{A}'$ via crossover and mutation
    \FOR{\textbf{each} $\mathbf{X} \in \mathcal{A} \cup \mathcal{A}'$}
        \STATE Evaluate $\mathbf{F}(\mathbf{X}) = [f_1, f_2, f_3]$ (using $H_j$) and constraint violations ($C_2$, $C_3$)
    \ENDFOR
    \STATE Perform non-dominated sorting and NSGA-II/NSGA-III survival selection
    \STATE Update termination statistics (objective-space, design-space, constraint-violation change)
    \STATE $g \leftarrow g+1$
\ENDWHILE
\STATE Extract Pareto set $\mathcal{P}$ from $\mathcal{A}$
\STATE $\mathbf{X}^* \leftarrow \arg\min_{\mathbf{X} \in \mathcal{P}} S(\mathbf{X})$ via AASF (Eq.~\eqref{eq:aasf})
\STATE Dispatch tasks according to $\mathbf{X}^*$ and update QNode queues
\end{algorithmic}
\end{algorithm}

\subsubsection{Schedule Encoding and Feasibility Handling}
We encode a schedule as a one-hot binary assignment matrix $\mathbf{X} \in \{0,1\}^{M\times N}$, which is directly compatible with pymoo’s binary MOEA operators and constraint interface \cite{pymoo}.
The unique-assignment constraint $C_2$ ($\sum_{j} x_{ij}=1,\ \forall i$) is enforced as an equality constraint, while the qubit-capacity constraint $C_3$ ($x_{ij}\, q_i^{r} \le q_j^{c}$) is enforced as an inequality constraint. Infeasible task-to-QNode pairs are discouraged by assigning a large penalty during fitness evaluation.

\subsubsection{Fitness Evaluation}
Given an assignment $\mathbf{a}$, the scheduler builds $\mathbf{X}$ by setting $x_{i,a_i}=1$ and evaluates the three-objective vector $\mathbf{F}(\mathbf{X})=[f_1,f_2,f_3]$ using the models in Section~II. The time objective uses the Heartbeat-based waiting-time estimate $H_j$, the error objective uses calibrated gate and readout error rates of the QNodes, and the stability objective uses the dispersion of completion times $L_j$.

\subsubsection{Dispatch and Queue Update}
After selecting $\mathbf{X}^*$, the Scheduler dispatches tasks to QNodes. Each QNode enqueues tasks, updates its next available time, and reports the updated availability and queue state in subsequent heartbeats; the resulting queueing delay is captured by $H_j$ and the accumulated busy time in $L_j(\mathbf{X})$.

\subsubsection{Fault Tolerance}
\noindent If a QNode becomes unavailable mid-batch, the Scheduler detects the failure via heartbeat timeout, marks the QNode offline, and re-queues any not-yet-started tasks for re-optimization in the next orchestration round; tasks that fail during execution are retried.

\section{Performance Evaluation}

\subsection{Environment Setup}

\subsubsection{Infrastructure and Software Stacks}
We implement and deploy QCOEM on Google Kubernetes Engine v1.33 and implement the scheduler and QNodes in Python v3.12, using Qiskit v2.0.1 for circuit transpilation and calibration-aware noise modeling, and pymoo\footnote{\url{https://pymoo.org/}} v0.6.1.5 for NSGA-II/NSGA-III optimization.

\subsubsection{Heterogeneous QNode Configuration}
We instantiate $N=5$ QNodes as Kubernetes pods with fixed resource limits. Each QNode emulates an IBM Quantum backend via a Qiskit \texttt{FakeBackendV2} calibration snapshot, yielding a pool of heterogeneous quantum computation resources.


\subsubsection{MOEA Hyperparameters}

We use pymoo's NSGA-II/NSGA-III implementations with population size $P=300$, up to $G_{\max}=1000$ generations, two-point crossover ($p_c=0.9$), and bitflip mutation. Fitness evaluations are parallelized with 512 threads and use \texttt{DefaultMultiObjectiveTermination} for convergence-based early stopping. AASF selection uses $\rho=10^{-4}$ and we report results across multiple random seeds.

\subsection{Workload Characteristics}
We create a workload from MQTBench \cite{quetschlich2023mqtbench} including representative hybrid algorithms (such as VQE and QAOA). To match the heterogeneous backend pool, we restrict tasks to 2-15 qubits and model arrivals as a Poisson process with rate $\lambda = 2.0$ tasks/s.

\subsection{Evaluation Metrics}
We report five metrics covering latency, computational quality, and operational stability.

\begin{itemize}
    \item \textit{Mean Task Completion Time (MCT)}: for each task $\tau_i$ assigned to QNode $q_j$, we compute waiting time from the heartbeat-predicted next-available time $A_j$ and execution time $t_{exec}(\tau_i,q_j)$, yielding $C_i=t_{wait,i}+t_{exec}(\tau_i,q_j)$ The MCT over $K$ completed tasks is:

    \begin{equation}
        MCT = \frac{1}{K} \sum_{i=1}^{K} C_i.
    \end{equation}

    \item \textit{Mean Fidelity}: the mean of the estimated fidelity $\mathcal{F}(\tau_i,q_j)$ (Section~II), reported as a counterpart of the error-rate objective $f_2$.

    \item \textit{Load Imbalance}: standard deviation of predicted QNode completion times $L_j$ across $N$ QNodes (lower is better).

    \item \textit{Rescheduling Rate}: fraction of tasks that require reassignment at dispatch due to infeasibility (e.g., violating qubit-capacity constraints or stale availability/queue state).

    \item \textit{Scheduling Overhead}: wall-clock time for the scheduler to produce an assignment per orchestration round.
\end{itemize}

\subsection{Performance Study}\label{AA}

\subsubsection{Baseline Performance Comparison}
\begin{figure}[!t]
    \centering
    \subfloat[Mean task completion time]{%
        \includegraphics[width=0.95\linewidth]{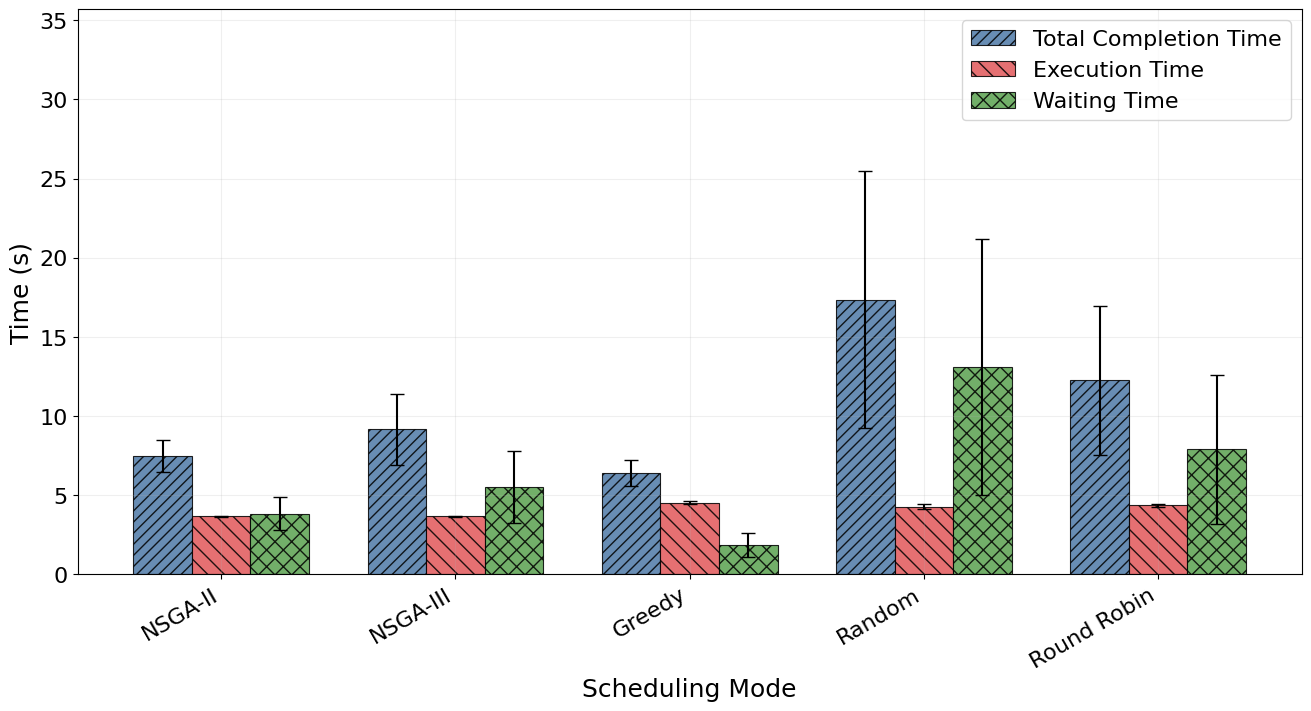}%
        \label{fig:sub_time}%
    }\\[1.0em]
    \subfloat[Mean fidelity and task rescheduling rate]{%
        \includegraphics[width=0.88\linewidth]{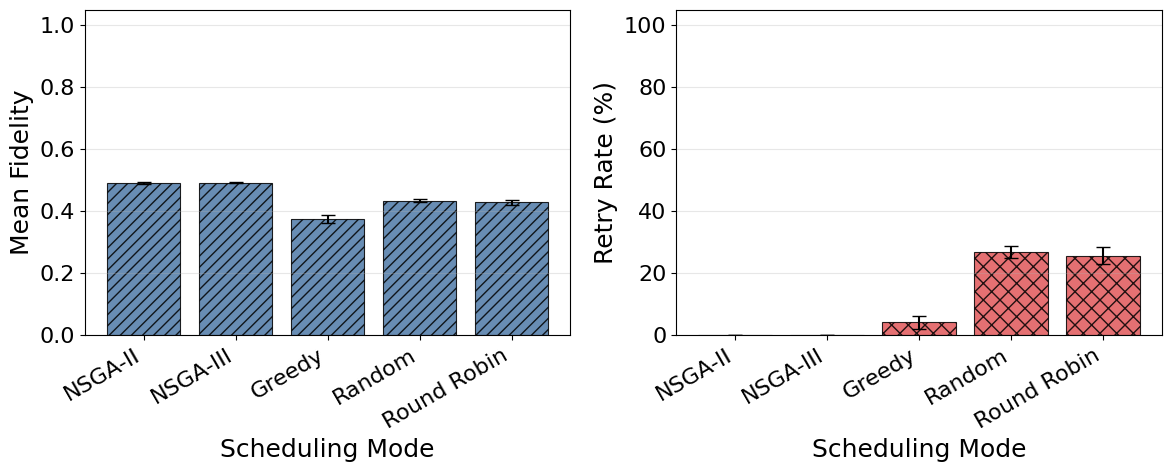}%
        \label{fig:sub_fidelity}%
    }\\[1.0em]
    \subfloat[Task distribution across QNodes]{
        \includegraphics[width=0.95\linewidth]{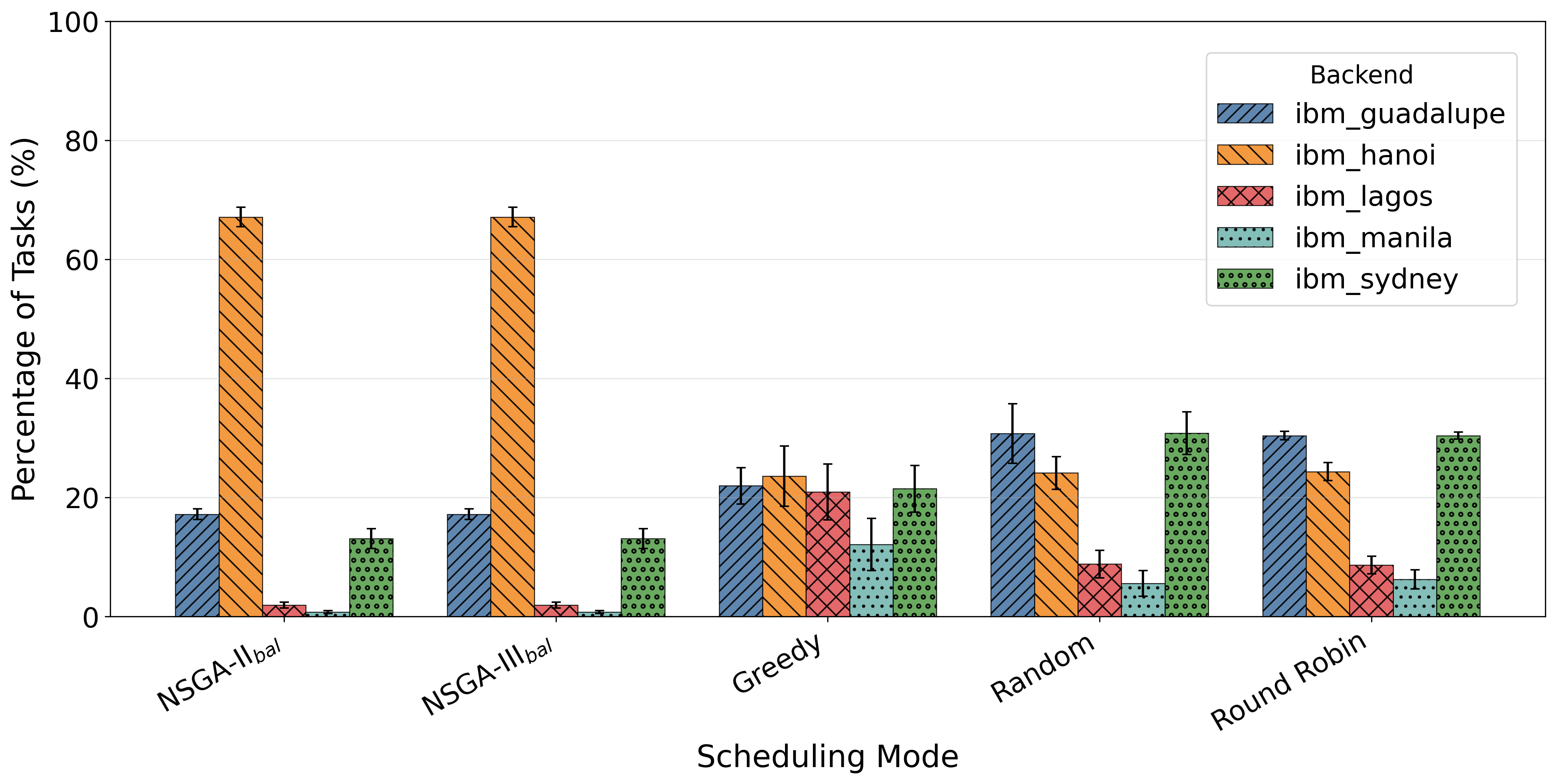}%
        \label{fig:sub_task_dist}%
    }
    \caption{Performance evaluation with baseline approaches: (a) time efficiency (MCT), (b) computational quality and resilience, and (c) task distribution across QNodes.}
    \label{fig:baseline_performance_combined}
\end{figure}

We evaluate the orchestration performance of the proposed Multi-Objective Evolutionary Algorithms (NSGA-II and NSGA-III) against three common baseline heuristics widely used in quantum cloud orchestrators: Greedy, Round-Robin, and Random \cite{nguyen2024drlq, li2024moirai, 10.1145/3712285.3759785} (see Fig.~\ref{fig:baseline_performance_combined}). These heuristics optimize throughput or latency in classical clusters but ignore quantum noise and hardware constraints. Comparing against them quantifies the performance loss from neglecting backend reliability and motivates a noise-aware framework for high-fidelity quantum workloads. For this comparison, MOEAs use AASF with a balanced preference vector $w=[0.35,0.45,0.20]$ for $[f_1,f_2,f_3]$, indicating a slight emphasis on fidelity while preserving load balancing. 

Noise-agnostic heuristics can minimize latency but often sacrifice fidelity and may trigger infeasible assignments. For instance, Greedy achieves a lower MCT but only reaches a mean fidelity of approximately $0.376$, while NSGA-II/NSGA-III raise the fidelity to about $0.49$ with zero retries, at the expense of a higher load imbalance to the QNode with better performance. Noise- and delay-aware multi-objective scheduling therefore improves execution quality and dispatch feasibility compared to classical heuristics, while preserving predictable load distribution.

\subsubsection{Sensitivity Analysis and Trade-offs}

\begin{figure}[t!]
    \centering
    \subfloat[Mean task completion time]{%
        \includegraphics[width=0.95\linewidth]{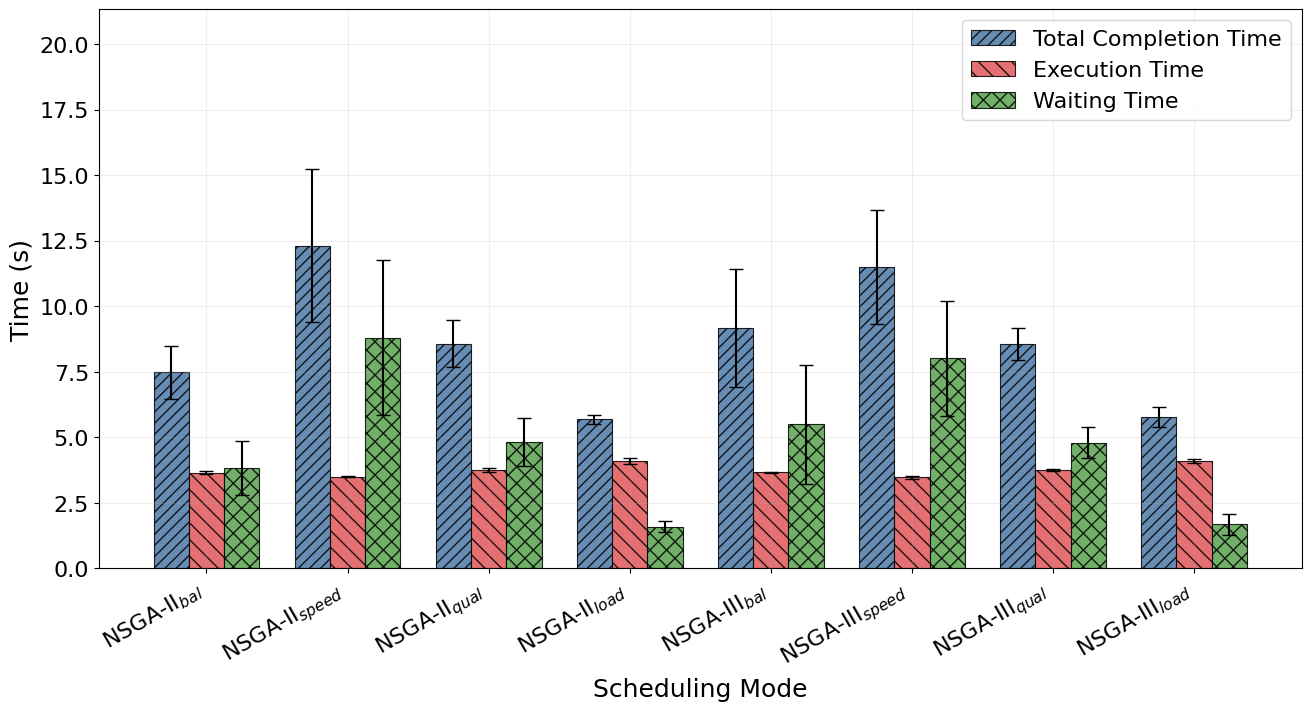}%
        \label{fig:time_metrics}%
    }\\[0.6em]
    \subfloat[Mean fidelity]{%
        \includegraphics[width=0.95\linewidth]{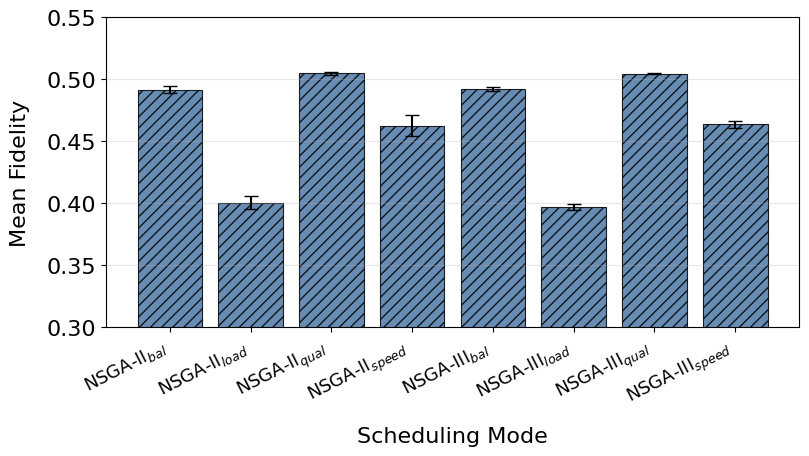}%
        \label{fig:mean_fidelity}%
    }
    \caption{Performance under different preference settings (bal: Balance, speed: Speed-Focus, qual: Quality-Focus, load: Load-Focus)}
    \label{fig:overall_results}
\end{figure}

While Fig.~\ref{fig:baseline_performance_combined} demonstrates the proposed framework's superiority over the baseline, practical quantum clouds must account for diverse user requirements and Service Level Agreements (SLAs). In this scenario, we evaluate the sensitivity of AASF-based decision making to different weight configurations $w=[w_{time}, w_{fid}, w_{load}]$. We consider four preference settings $w$: \textit{Balanced} ($[0.35,0.45,0.20]$), \textit{Speed-Focus} ($[0.90,0.05,0.05]$), \textit{Quality-Focus} ($[0.05,0.90,0.05]$), and \textit{Load-Focus} ($[0.05,0.05,0.90]$), with a minimum weight of $0.05$ on non-primary objectives to avoid degenerate scalarization. Fig.~\ref{fig:overall_results} illustrates the time-fidelity trade-offs.

\textit{a) Load-Focus} reduces total completion time by mitigating queueing delays. In heterogeneous QNode pools, optimizing only for short execution time concentrates tasks on low-latency QNodes, increasing queueing delays and degrading end-to-end completion time. In contrast, \textit{Load-Focus} promotes a more uniform task distribution, yielding the lowest total completion time (e.g., NSGA-II at approximately $5.68$s).

\textit{b) Quality-Focus} improves outcome fidelity but increases latency and load imbalance. Compared to the \textit{Balanced} configuration (NSGA-II fidelity approximately $0.491$), it raises mean fidelity to approximately $0.505$, while extending total completion time from $7.49$s to $8.58$s and increasing load imbalance from $493.54$s to $643.31$s.

\textit{c) Speed-Focus} minimizes execution time but can degrade end-to-end latency due to queueing and hotspotting. Relative to \textit{Balanced}, \textit{Speed-Focus} achieves the shortest execution time (NSGA-II at about $3.50$s versus \ $3.66$s) but incurs substantially higher waiting times (about $8.81$s versus \ $3.83$s), resulting in a higher overall completion time (roughly $12.31$s versus\ $7.49$s). This aggressive routing further lowers mean fidelity, approximately $0.462$ versus \ $0.491$, and markedly increases load imbalance at about $1199.33$s compared to \ $493.54$s.

\textit{d) Balanced} constitutes a practical operating point, while NSGA-II demonstrates greater robustness. The \textit{Balanced} configuration yields a favorable trade-off between completion time and fidelity (Fig.~\ref{fig:overall_results}). Across repeated runs, NSGA-II exhibits lower variance than NSGA-III (e.g., under \textit{Balanced}, NSGA-II at approximately $7.49$s versus \ NSGA-III at $9.17$s), indicating that NSGA-II is a more reliable default optimization engine in this three-objective setting. Overall, preference weights materially change routing behavior through queueing dynamics; explicitly modeling load imbalance helps avoid hotspotting when optimizing for speed or fidelity.

\subsubsection{Scalability and Overhead}

We measure scheduling overhead as the wall-clock time to obtain a dispatchable assignment for batch sizes $M\in\{5,10,20,50\}$, excluding hardware execution. Heuristics form a lower bound (Random/Round Robin at about $ 10^{-4}$s), while Greedy grows roughly linearly from $0.78$s ($M=5$) to $7.64$s ($M=50$). MOEAs incur higher control-plane cost from evolutionary search; at $M=50$, NSGA-II and NSGA-III take about $94.58$s and $84.80$s, with NSGA-II converging in $12.4$ times the scheduling time of Greedy at this batch size. Nevertheless, in QaaS settings where QPU queueing dominates end-to-end latency, MOEA-based routing is preferable for moderate-to-large batches (or longer scheduling intervals), where gains in feasibility and fidelity can outweigh the added control-plane latency.

\section{Related Work}

Quantum cloud research has yielded middleware, simulators, and schedulers; however, the problem of real-time, noise-aware orchestration across heterogeneous NISQ backends remains largely unresolved. Cloud-native middleware platforms can integrate QPUs into containerized infrastructures such as Kubernetes, yet they typically model backends as static resources and provide only limited capabilities for noise- and topology-aware routing \cite{stirbu2024qubernetes, faro2023middleware, saurabh2023conceptual}. Simulation, emulation, and digital-twin frameworks support scalable benchmarking and increasingly incorporate calibration-informed noise models, but they are predominantly employed as offline experimentation environments rather than as production-grade orchestration engines \cite{nguyen_iquantumtoolkitmodeling_2024, 10.1145/3726301.3732296, 11000150}. Existing schedulers encompass heuristic and utility-based routing as well as deep reinforcement learning and evolutionary strategies; nevertheless, many approaches either optimize exclusively for latency \cite{nguyen2024qfaas, ravi2021adaptive, nguyen2024drlq} and fidelity or aggregate multiple objectives into fixed weighted sums, which can neglect solutions on non-convex Pareto fronts and intensify load imbalance \cite{10.1145/3754598.3754641, li2025qusplit}. Notably, Qonductor \cite{10.1145/3712285.3759785} applies NSGA-II to quantum scheduling, and QFOR \cite{nguyen2025qfor} uses deep reinforcement learning for fidelity-aware orchestration; however, neither jointly models load imbalance nor applies AASF-based preference selection, which distinguishes QCOEM. In summary, prior work still lacks an end-to-end, cloud-native orchestration workflow that jointly models noise, queueing dynamics, and load imbalance and supports preference-driven multi-objective decision-making, which our QCOEM framework is proposed to address.

\section{Conclusions and Future Work}
Integrating quantum computing into cloud-native infrastructures is challenging due to the heterogeneity and noise sensitivity of NISQ hardware. Existing schedulers rely on noise-agnostic heuristics or static scalarization, lacking the fine-grained control needed for resource allocation in non-convex objective spaces. Our QCOEM framework formulates quantum task assignment as a multi-objective optimization problem, minimizing completion time, execution error rate, and load imbalance, and solves it by leveraging NSGA-II and NSGA-III approaches. Using an Augmented Achievement Scalarization Function, we perform preference-based selection from the MOEA-generated candidate set to obtain a dispatchable schedule weighted by user preferences. In a heterogeneous quantum cloud environment, QCOEM eliminates task rescheduling and improves mean fidelity by approximately 30\% over noise-agnostic baselines while keeping scheduling overhead bounded. These results suggest that noise-aware scheduling can significantly improve fidelity and cluster stability, forming a basis for more predictable, high-performance quantum cloud services. To enhance the comprehensiveness of our study, we will expand the evaluation by benchmarking our approach against a broader set of state-of-the-art techniques for quantum cloud scheduling, such as deep reinforcement learning and by validating QCOEM under more realistic conditions, including experiments on real quantum hardware and simulated environments to better characterise latency-fidelity-cost trade-offs. These future directions aim to advance our orchestration into a robust, scalable, and noise-aware framework for quantum cloud computing.

\section*{Acknowledgment}
This research was supported by the VNUHCM-University of Information Technology's Scientific Research Support Fund. The Overleaf AI Assistant was used to correct grammar and spelling during manuscript preparation. All intellectual content and analysis are the sole work of the authors, who take full responsibility for the accuracy of this work.

\bibliographystyle{IEEEtran}

\bibliography{references}

\end{document}